\documentclass[utf8]{FrontiersinHarvard} 

\usepackage{url,hyperref,lineno,microtype,subcaption}
\usepackage[onehalfspacing]{setspace}

\def\keyFont{\fontsize{8}{11}\helveticabold }
\def\firstAuthorLast{Madsen {et~al.}} 
\def\Authors{
Kristin Madsen\,$^{1,*}$, 
Javier A. Garc\'ia\,$^{1,5}$,  
Daniel Stern\,$^{2}$,

Rashied Armini$^{2}$,
Stefano Basso$^{3}$,
Diogo Coutinho$^{4}$,
Brian W. Grefenstette$^{5}$,
Steven Kenyon$^{1}$,
Alberto Moretti$^{3}$,
Patrick Morrisey$^{2}$ 
Kirpal Nandra$^{4}$,
Giovanni Pareschi$^{3}$,
Peter Predehl$^{4}$,
Arne Rau$^{4}$,
Daniele Spiga$^{3}$,
Jörn Willms$^{6}$,
William W. Zhang$^{1}$
}

\def\Address{$^{1}$X-ray Astrophysics Laboratory, NASA Goddard Space Flight Center, Greenbelt, MD 20771, USA\\
$^{2}$Jet Propulsion Laboratory, California Institute of Technology, Pasadena, CA 91109, USA\\
$^{3}$INAF-Osservatorio Astronomico di Brera, Via E. Bianchi 46, Merate (LC), Italy\\
$^{4}$Max Planck Institute for Extraterrestrial Physics, Giessenbachstrasse 1, 85748 Garching, Germany\\
$^{5}$Cahill Center for Astrophysics, California Institute of Technology, 1216 E California Blvd., Pasadena, CA 91125, USA\\
$^{6}$Friedrich-Alexander-Universität, Erlangen-Nürnberg, Erlangen, Bayern, Germany}
\def\corrAuthor{Kristin K. Madsen}

\def\corrEmail{kristin.c.madsen@nasa.gov}

\usepackage[dvipsnames]{xcolor}

\newcommand{\nustar}{\textit{{NuSTAR}}}

\newcommand{\xmm}{\textit{XMM-Newton}}

\newcommand{\as}{$''$}

\begin{document}
\onecolumn
\firstpage{1}

\title {The High Energy X-ray Probe (HEX-P): Instrument and Mission Profile\\ }

\author[\firstAuthorLast ]{\Authors} 
\address{} 
\correspondance{} 

\extraAuth{}

\maketitle

\begin{abstract}
The High Energy X-ray Probe (HEX-P) is a proposed NASA probe-class mission that combines the power of high angular resolution with a broad X-ray bandpass to provide the necessary leap in capabilities to address the important astrophysical questions of the next decade. HEX-P achieves breakthrough performance by combining technologies developed by experienced international partners. HEX-P will be launched into L1 to enable high observing efficiency. To meet the science goals, the payload consists of a suite of co-aligned X-ray telescopes designed to cover the 0.2 -- 80 keV bandpass. The High Energy Telescope (HET) has an effective bandpass of 2 -- 80 keV, and the Low Energy Telescope (LET) has an effective bandpass of 0.2 -- 20 keV. The combination of bandpass and high observing efficiency delivers a powerful platform for broad science to serve a wide community. The baseline mission is five years, with 30\% of the observing time dedicated to the PI-led program and 70\% to a General Observer (GO) program. The GO program will be executed along with the PI-led program.


\tiny
 \keyFont{ \section{Keywords:} X-ray, HEX-P, mission} 
\end{abstract}

\section{Introduction}
The current landscape for large X-ray missions is dominated by oversubscribed observatories far into their extended operation phase. The next planned flagship-class X-ray observatory is the ESA Athena mission, currently scheduled to launch no earlier than 2037. With the current outlook, the science community faces a non-negligible risk of being without a sensitive X-ray GO observatory in the 2030s, which is particularly alarming given the extensive number of time domain and multi-messenger astrophysics (TDAMM) facilities coming on line over the next decade and the critical role high-energy observations play in understanding the physics of explosive and energetic phenomena \citep[e.g.,][]{Brightman2023}. HEX-P is designed to bridge this gap, and complement Athena once it launches, with instrumentation that will deliver capabilities to advance our understanding of the X-ray Universe across a breadth of topics.

The \textit{Nuclear Spectroscopic Telescope Array} (\nustar) \citep{Harrison2013}, a NASA Small Explorer mission,  opened the high-energy ($>$~10~keV) X-ray band by establishing the power of focused broad-band X-ray spectroscopy (3 -- 80~keV). HEX-P builds on this legacy with a mission concept submitted to the 2023 NASA Astrophysics Probe Explorer (APEX) call, which was directed by the 2020 Decadal Survey on Astronomy and Astrophysics\footnote{https://www.nationalacademies.org/our-work/decadal-survey-on-astronomy-and-astrophysics-2020-astro2020}. HEX-P combines the power of high angular resolution with broad bandpass coverage to provide the necessary leap in capabilities to address important astrophysical questions of the next decade. These questions were formulated in the 2020 Decadal Survey and motivate new observational capabilities across the electromagnetic and multi-messenger regimes to resolve the rich workings of planets, stars, and galaxies on all scales. HEX-P fills the important gap between soft X-rays and $\gamma$-rays where compact objects, from stellar remnants (neutron stars and black holes) up to supermassive black holes, emit vast amounts of energy through the process of accretion. 

\begin{figure}
    \centering
    \includegraphics[width=0.55\textwidth]{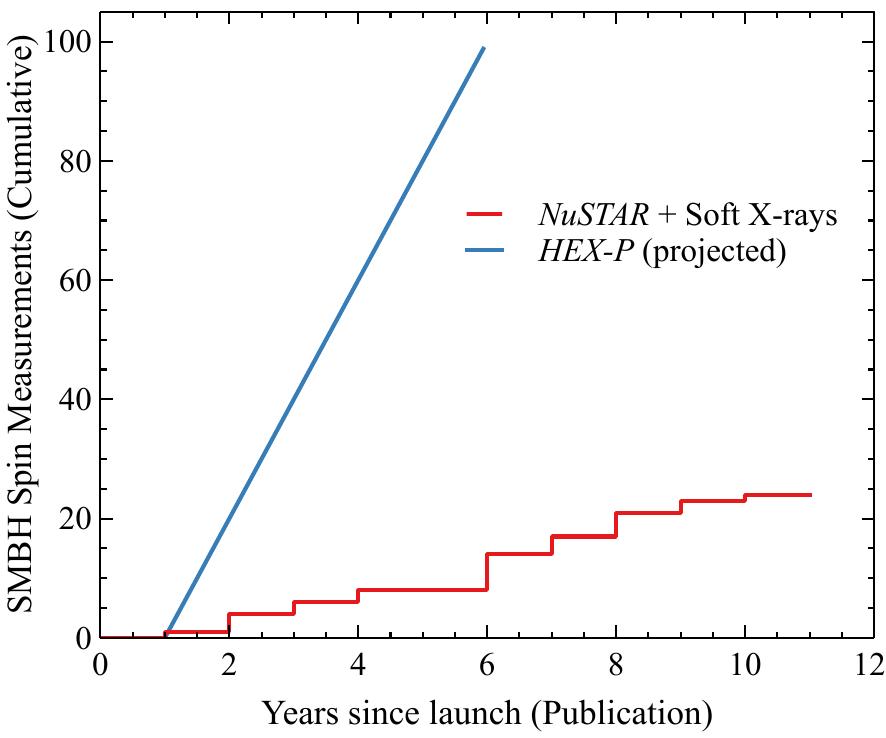}
    \caption{Time to obtain 100 supermassive black hole (SMBH) spin measurements with HEX-P compared to \nustar\ observations coordinated with soft X-ray observatories (primarily \xmm). The number of spin measurements obtained with \nustar\ are collected from published papers and demonstrates that at the current rate, which is limited by the coordinated observing time, it would take about 30 years to obtain 100 spin measurements through GO programs.}
    \label{fig:spin}
\end{figure}

HEX-P will be launched into L1, which provides $\ge 90\%$ observing efficiency, a large instantaneous field-of-regard, and the capability to do long, uninterrupted observations. HEX-P's combination of bandpass and high observing efficiency delivers a powerful General Observer (GO) platform for a broad range of science that services a wide community base. The advantage of combining the soft and hard X-ray bandpasses in a single observatory allows for the exploration of science that is currently challenging due to the necessity of coordinating multiple observatories across several agencies. \nustar, the only available focusing hard X-ray satellite currently in orbit, spends $>$~50\% of its observing time coordinating with other observatories. This is limited by the time each agency is willing to share for coordination and is becoming increasingly challenging as the desire for coordinated hard and soft X-ray observations has risen, e.g., with the recent launches of NASA's \textit{Imaging X-ray Polarimetry Explorer} (\textit{IXPE}) and JAXA's \textit{X-ray Imaging and Spectroscopy Mission} (\textit{XRISM}).  The demand for simultaneous observations is oversubscribed by a factor of five and increases each cycle. As an example, HEX-P's plan to measure the spin distribution of 100 AGN \citep{Piotrowska2023} would take approximately 30~years with the current level of coordination between observatories in their GO programs (Figure~\ref{fig:spin}).

\begin{figure}
    \centering
    \includegraphics[width=0.95\textwidth]{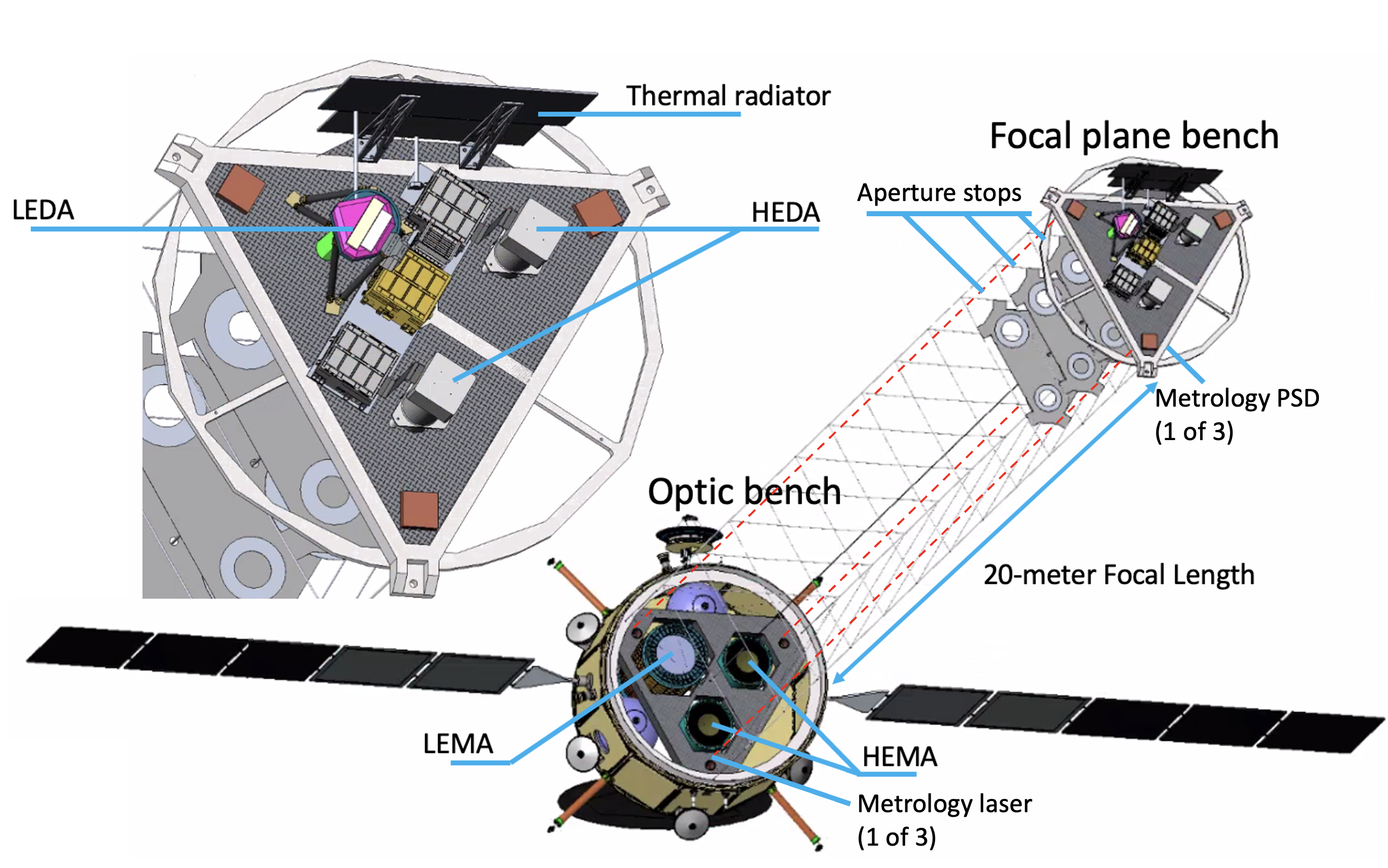}
    \caption{Three Wolter-I mirror assemblies are mounted rigidly onto the spacecraft structure and connected to a composite focal plane bench by a 20-m long, three-sided, coilable boom. A metrology system comprised of three laser-detector pairs provides real-time tracking of displacements between the benches that are used to reconstruct images from time-resolved data during post-processing. Aperture stops interior to the boom provide background mitigation from stray-light and the diffuse cosmic X-ray background.}
    \label{fig:observatory}
\end{figure}

\section{Observatory and Mission Overview}
The HEX-P payload consists of a suite of three co-aligned X-ray telescopes designed to cover the 0.2 -- 80~keV bandpass. The High Energy Telescope (HET) has an effective bandpass of 2--80~keV, and is composed of two High Energy Mirror Assemblies (HEMAs), which utilize electroformed Ni shells coated with Pt/C and W/Si multilayer coatings for high energy response. These focus onto two High Energy Detector Assemblies (HEDAs), which are photon counting CdZnTe (CZT) hybrid detectors based on the \nustar\ design. The Low Energy Telescope (LET) has an effective bandpass of 0.2--20~keV and is composed of a Low Energy Mirror Assembly (LEMA) optimized for improved spatial resolution that capitalizes on Si segmented mirror technology developed at the NASA Goddard Space Flight Center (GSFC). The LEMA focuses onto the Low Energy Detector Assembly (LEDA), which utilizes a Depleted P-Channel Field Effect Transistor (DEPFET).

The HEX-P instrumentation is shown in Figure~\ref{fig:observatory}. The three Wolter type I (Wolter-I) telescope mirror assemblies are mounted to a Northrop Grumman three-axis stabilized spacecraft bus that serves as the base for a three-sided, coilable boom, extended after launch to place the detector bench at a focal length of 20 m. The coilable boom is enclosed in a thermal Kapton sock to suppress stray-light and bremsstrahlung activation of the boom components by solar X-ray flux. The boom deploys with an array of three aperture stops to mitigate stray-light from sources outside the field-of-view (FoV) and the cosmic X-ray background (CXB). 

A metrology system (MET), comprised of three parallel diode lasers mounted to the optics bench and co-aligned with the telescopes, forms a reference for three position sensitive detectors (PSDs) mounted at focal plane bench. The MET tracks motions of the boom during observations, while three bus-mounted star trackers provide an absolute pointing reference that is combined with the laser metrology to reconstruct data on the ground.

The baseline mission is five years with 30\% of the observing time dedicated to the PI-led program and 70\% to a GO program. The GO program will run alongside the PI-led program, and all data will be made available without proprietary time.

\section{Payload Instruments}
\subsection{The High Energy Telescope: HET}
The HET consists of the two HEMA optics, which are co-aligned and observe the same field. These two mirror assemblies are identical and coated with Pt/C and W/Si multilayers to achieve the broad bandpass response. The two HEMAs are matched to the HEDAs, which are photon counting CZT hybrid detectors based on the \nustar\ design built by Caltech. The HET has a FoV of 13.7\as$\times$13.7\as, driven by the detector size, and the mirrors are expected to provide a half-power diameter (HPD) angular resolution of 10\as\ at 6 keV and 23\as\ at 60 keV. 

The performance presented in this paper corresponds to the instrument specifications that are summarized in Table \ref{tab:hexp}, corresponding to the version 7 (v07) instrument design, which was distributed in April 2023. For the submission of the proposal in November 2023, the number of HEMA shells was reduced from 60 to 57, resulting in a 12\% loss of HET effective area.

\begin{table}[]
    \centering
    \caption{HEX-P instrument performance (v07).}
    \begin{tabular}{l|c|c}
        \hline
                             & LET                 & HET                                   \\
        \hline
        \# optic;   detectors & 1                   & 2                                     \\
        Focal Length          & 20 m                & 20 m                                  \\
        Bandpass              & 0.2 – 20 keV        & 2 – 80 keV                              \\
        \hline
        Detectors             & LEDA                & HEDA                                  \\
        \hline
        Field of view         & 11.3\as $\times$ 11.3\as       & 13.7\as $\times$ 13.7\as                         \\
        Detector sensor       & DEPFET Si           & CnZnTe (CZT) pixel hybrid             \\
        Sensor format         & 512 $\times$ 512 pixels    & 32 $\times$ 32 pixels                        \\
        \# Sensors            & 1                   & 16 (4 x 4)                            \\
        Sensor   thickness    & 450 µm              & 3 mm                                  \\
        Pixel pitch           & 130 µm              & 604.8 µm                              \\
        Time resolution       & full-frame: 2ms (w64: 250 $\mu$s) & 1 $\mu$s                            \\
        \hline
        Mirrors               & LEMA                & HEMA                                  \\
        \hline
        Mirror   technology   & Monocrystalline Si  & Ni electroformed                      \\
        \# Shells             & 24 [30]             & 60 [57]                               \\
        Half-power diameter   & 3.5\as\ [2.9\as]  & 10\as, 17\as, 23\as @ 10, 30, 60 keV\\
        Maximum radius        & 350 mm              & 250 mm                                \\
        Shell length          & 800 mm              & 660 mm \\
        Shell thickness       & 0.8 mm              & 0.32 – 0.55 mm                          \\
        Mirror coating        & Ir with C overcoat  & Pt/C \& W/Si graded multilayers       \\
        \hline
        \multicolumn{3}{l}{[*] Changes to the baseline not included in the instrument performance presented in this paper.}
    \end{tabular}
    \label{tab:hexp}
\end{table}

\subsubsection{The High Energy Mirror Assembly: HEMA}
The two HEMAs will be made from complete shells, replicated from high-accuracy mandrels by Media Lario (Bosisio Parini, Italy) by means of Ni electroforming \citep{Vernani2011}. This technology has been successfully used for the Ni gold-coated X-ray mirrors of \textit{BeppoSAX} (launched in 1990), \textit{Swift} (1999), \xmm\, (1999), \textit{SRG}/eROSITA (2017), and is scheduled to be launched in December 2023 on the \textit{Einstein Probe} \citep{Vernani2022}.

\begin{figure}
    \centering
    \includegraphics[width= 0.5\textwidth]{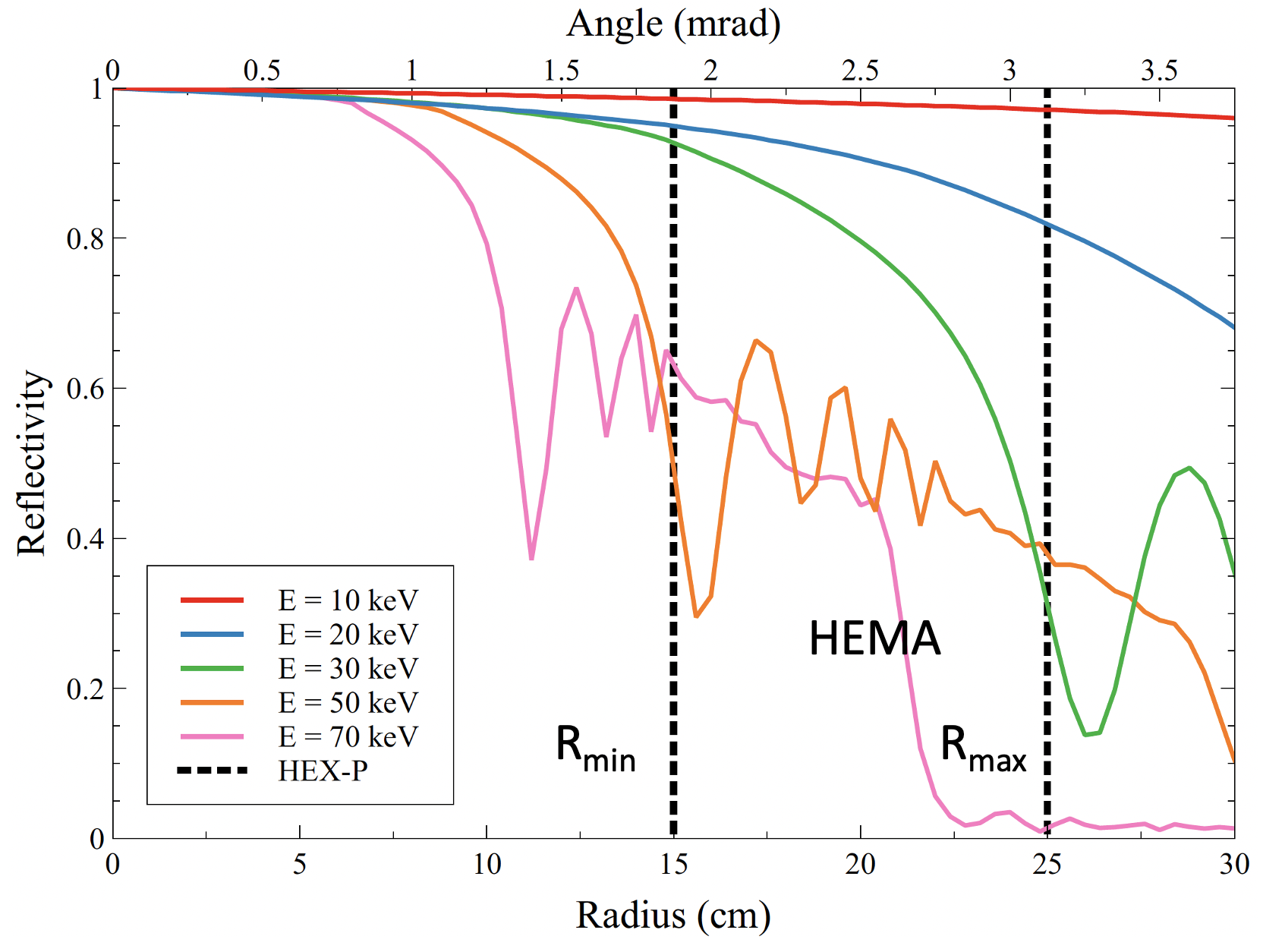}
    \caption{Reflectivity curve of a Pt/C multilayer coating as a function of energy.}
    \label{fig:angVr}
\end{figure}

Each HEMA has 60 (v07) nested Wolter-I paraboloid-hyperboloid mirror shells. The primary design constraint requires that the grazing incidence angle be less than 3.2~mrad.  For a focal length of 20~m, this corresponds to an outermost diameter at the parabola-hyperbola intersection plane of approximately 500 mm. Figure \ref{fig:angVr} shows the reflectivity of a Pt/C multilayer coating optimized for the bandpass, demonstating that at these angles the coatings have $>$50\% efficiency for energies up to 50 keV. Long mirrors provide more effective area per shell, and for this reason the HEMA shells are 660~mm tall, constrained by the maximum shell length that can be produced using the current Media Lario facilities.

The multilayers for the HEMA are optimized for bandpass and performance in a well-established process \citep{Madsen2004,Madsen2022}. The number of bilayers is $<100$ and the maximum bilayer thickness $<30$~ nm, making them both simpler and thinner than was done for \nustar ~\citep{Madsen2009}. The W K-edge is at 69.5~keV and the Pt K-edge is at 78.4~keV; beyond these energies, the respective coating efficiencies drop precipitously. Because the W/Si coating is smoother and easier to deposit, the HEMAs will have shells coated with W/Si at the outermost radii where the coatings only reflect $<70$~keV. The innermost shells, which can achieve 70-80 keV, will be coated with Pt/C.

On the basis of the previous experience gained with Ni electroformed shells, in particular eROSITA, an angular resolution of HPD $\le$ 10\as\ at energies $< 10$~keV is anticipated. The HPD trend at higher energies is dominated by the power spectral density (PSD) of the shells surface roughness, which is responsible for the X-ray scattering. The HEX-P HPD is estimated from the PSD of the eROSITA mirror modules. Extrapolating eROSITA in-flight performance to the energy range and angles of HEX-P, the HPD angular resolution is anticipated to be $<$ 25\as\ for energies $\le$ 80~keV. A sample of HPDs calculated for selected energies is shown in Table~\ref{tab:hexp}, and a more detailed description of the HEX-P mirror design can be found in \cite{Madsen2023_1}.

\subsubsection{The High Energy Detector Assembly: HEDA}
The design of the HEDA is adapted from flight-proven hard X-ray detectors flown on \nustar\ with over a decade of successful operation in space \citep{Harrison2013}. These hybrid detectors are composed of a monolithic crystal of CZT flip-chip bonded to an Application Specific Integrated Circuit (ASIC) \citep{Chen2004}. To improve the high-energy quantum efficiency (QE), the HEDA uses 3-mm thick CZT rather than the 2-mm thick CZT used on \nustar. The dimensions of the individual hybrid detectors remain the same, and due to the longer focal length, each HEDA will use an array of $4\times 4$ detectors compared to the \nustar\ $2\times 2$ layout. This conserves the FoV to be 13.7\as$\times$13.7\as, as it is for \nustar. 

The energy resolution is comparable to \nustar, with a 500 eV full-width at half-maximum (FWHM) energy resolution below 10 keV,  increasing to 850~eV at 60~keV. Improvements to the digital processing and analog-to-digital conversion rates increase the overall throughput of the system by up to 1000 cps from 400 cps in \nustar. The detectors can handle an incidence countrate of 10,000 cps before pile-up becomes an issue. Like \nustar, the HEDA is surrounded by an anti-coincidence shield to remove the background due to charged particles.

\subsubsection{HET Background}
\begin{figure}
    \centering
    \includegraphics[width= 0.45\textwidth]{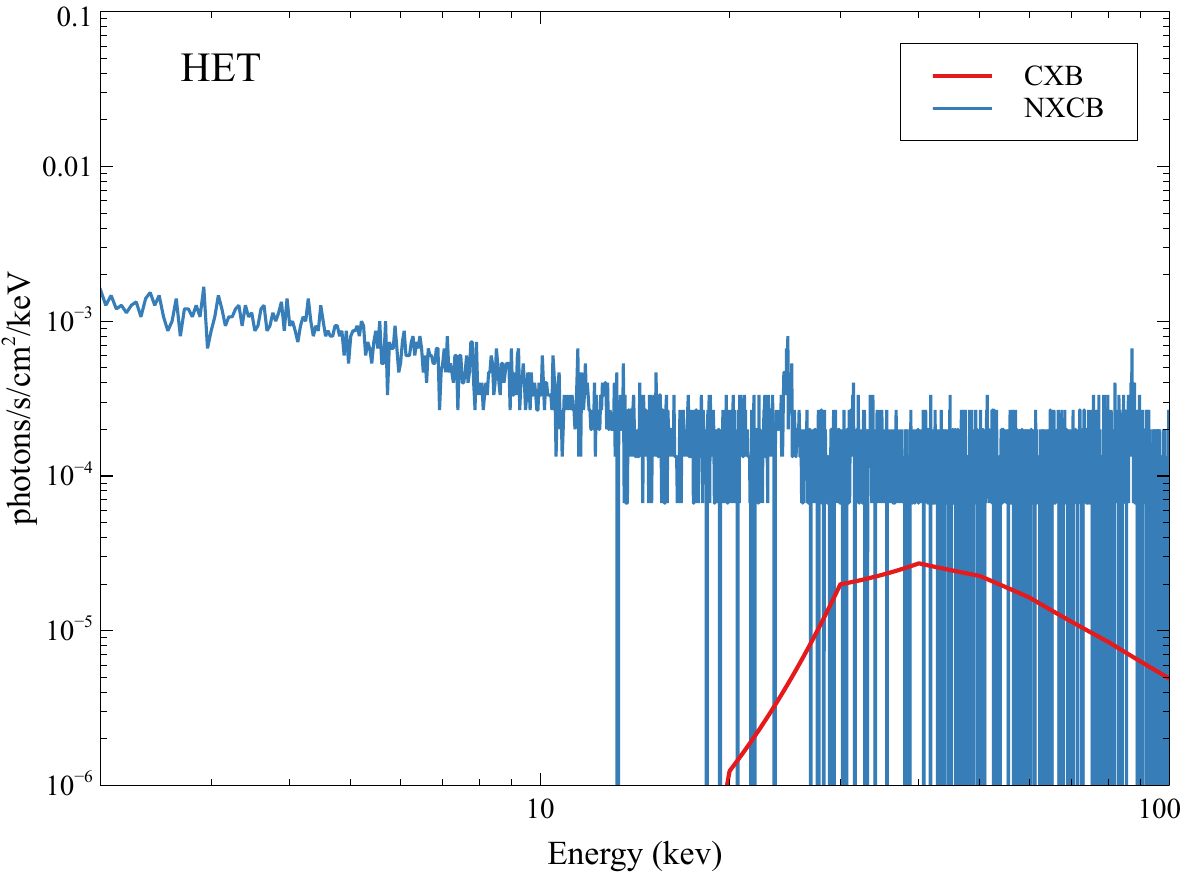}
    \includegraphics[width= 0.45\textwidth]{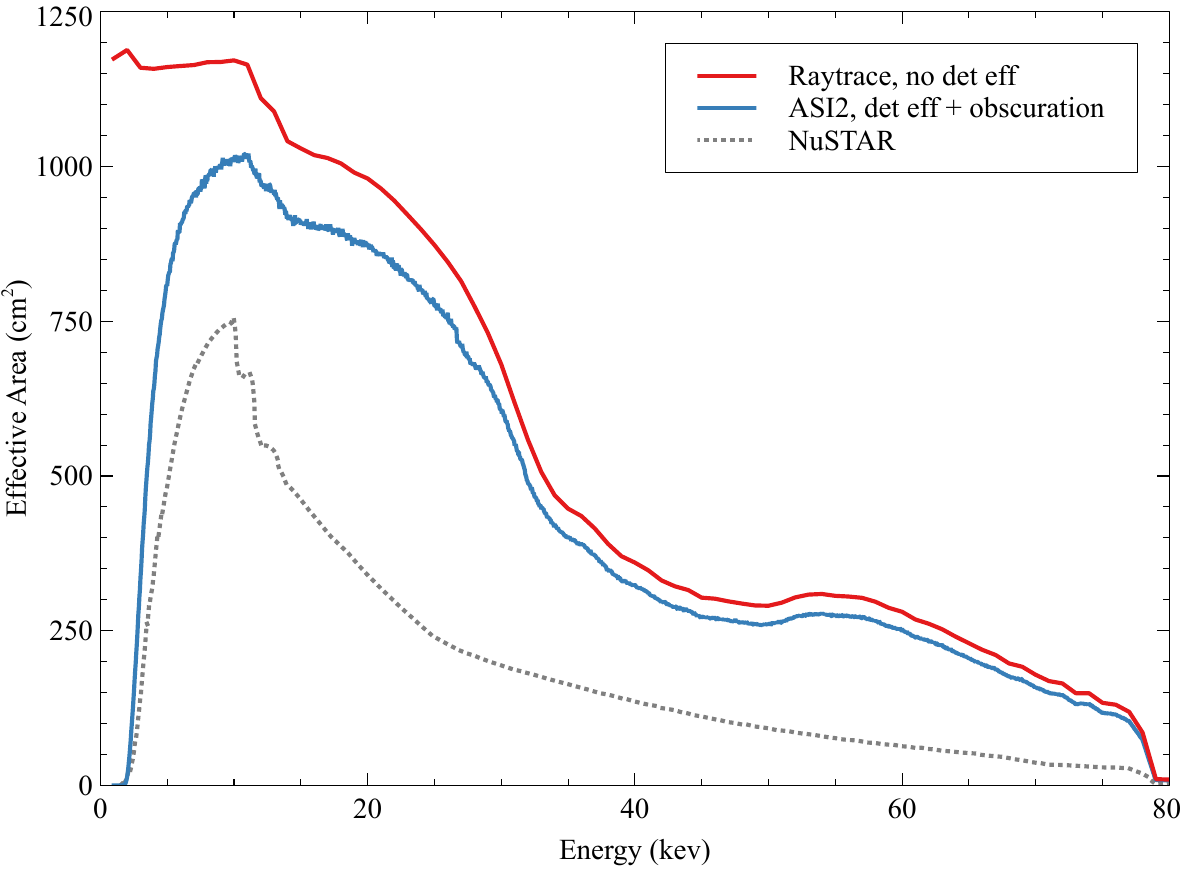}
    \caption{\textbf{Left}: The HET instrument background separated into the CXB and NCXB. \textbf{Right}: The effective area of the LET, with and without the detector efficiency and other absorbing elements in the X-ray path.}
    \label{fig:het_rsp}
\end{figure}

Due to the similarities between the \nustar\ focal plane module (FPM) and the HEDA, a complete mass model for a \nustar\ FPM was used to estimate the background contribution from the charged particle flux at L1 using GEANT4 modeling \citep{Madsen2023_2}. The baseline anti-coincidence shield is BiGeO (BGO) instead of the CsI used on \nustar\ in order to eliminate the Cs and I lines between 23-28~keV. This will instead produce a Ge K$\alpha$ lines at 9-11~keV and Bi K$\alpha$ lines at 74-77~keV. For the purpose of comparing to \nustar, simulations were performed for CsI and the results were used to scale the \nustar\ background. The background was fit with an empirical model, and the Cs and I lines were removed before producing the background instrument file for simulations. The HET background contribution from the CXB and the charged particle, non X-ray background (NCXB) is shown in Figure~\ref{fig:het_rsp} (left).

\subsubsection{HET Effective Area}
The effective area of the HEMA was derived from a full ray-trace, including all known effects of obscuration, surface mirco-roughness, and the mirror reflection from multilayers. The QE and redistribution matrix of the HEDA was simulated using the charge transport and detector effects modeling code developed for \nustar\ and validated against flight data. Figure \ref{fig:het_rsp} (right) shows the current best estimate of the effective area of HEMA folded through the HEDA detector redistribution function. 

\subsection{The Low Energy Telescope: LET}
The LET is composed of one LEMA mirror, which is optimized for improved spatial resolution up to 20~keV and capitalizes on the Si segmented mirror technology developed at the NASA GSFC. At the focus of the LEMA is the LEDA, which utilizes the DEPFET technology developed by the Max Planck Institute for Extraterrestrial Physics (MPE; Garching, Germany) for the Wide Field Imager (WFI) \citep{Meidinger2020} on the ESA L-class Athena mission \citep{Ayre2018,Bavdaz2021}. The FoV of the LET is 11.3\as$\times$11.3\as, and the anticipated angular resolution of the HEMA is projected to be HPD $\le$3\as. 

The performance presented in this paper corresponds to the instrument specifications that are summarized in Table \ref{tab:hexp} and corresponds to the `v07's estimated performance that were distributed in April 2023. For the submitted proposal, the number of shells of the LEMA were increased from from 24 to 30, resulting in a 35\% increase in effective area. In addition, the current best estimate of the LET HPD improved from 3.5\as\ to 2.9\as.

\subsubsection{The Low Energy Mirror Assembly: LEMA}
The LEMA uses the single crystal Si mirror technology that has been under development at GSFC by the Next Generation X-ray Optics team since 2010. It consists of semiconductor-grade monocrystalline segments, approximately 100 $\times$ 100~mm$^2$ and 0.8~mm in thickness, that are cut out of a block of Si and polished to an arcsecond figure using lapping and polishing techniques, including ion-beam figuring \citep{Riveros2022}.

The optical design and curvature of the primary and secondary mirror follow a variant of the Wolter-I design that employs a hyperbolic-hyperbolic prescription \citep{Harvey2001,Saha2022}. In this design, the Wolter-I parabolic primary is replaced with a hyperbolic curvature, which provides superior off-axis angular performance over a conventional Wolter-I design. The LEMA is a segmented mirror with the sub-components collected into 18 mirror modules that are arranged in a ring. Due to the long focal length, the mirror modules are required to be 800~mm in length, and each mirror module contains 8 (axial) and 24 (radial) mirror segments (for the v07 design).

\begin{figure}
    \centering
    \includegraphics[width=0.45\textwidth]{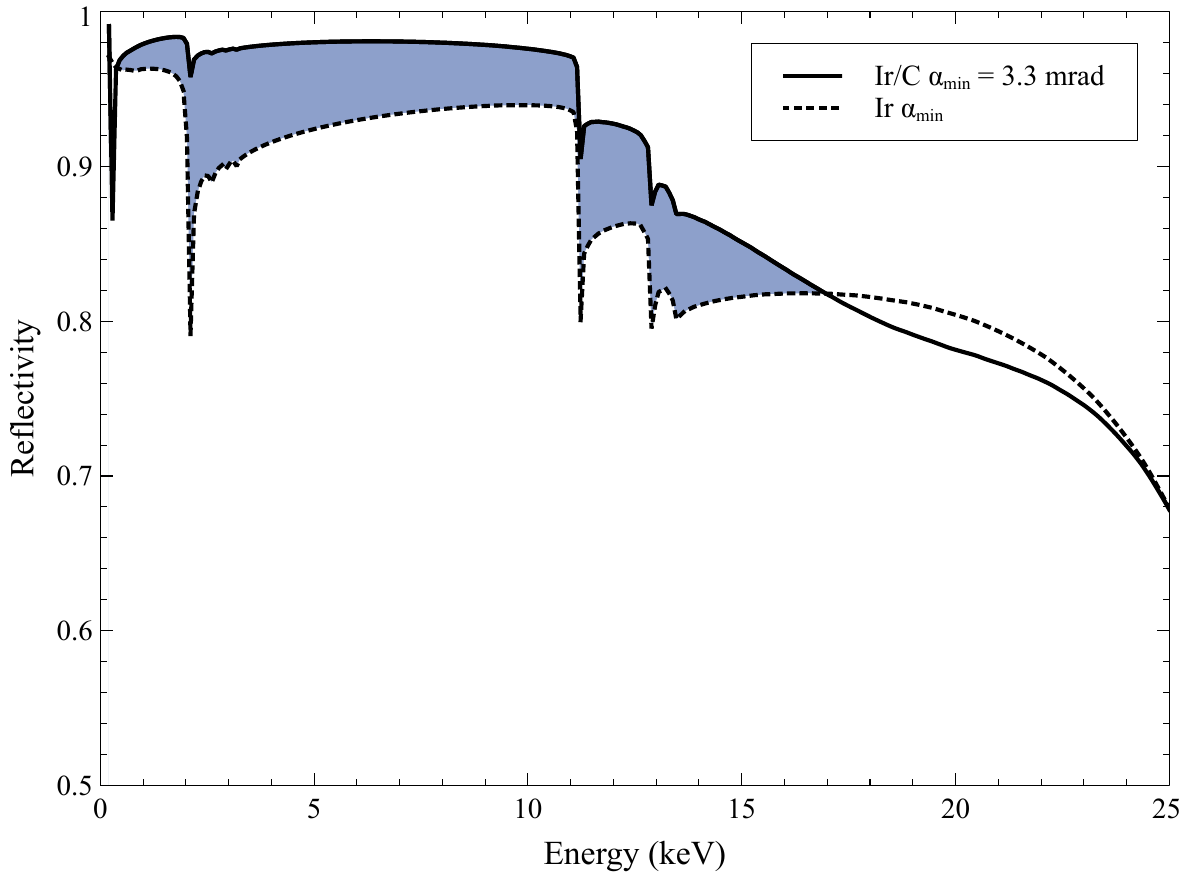}
    \includegraphics[width=0.45\textwidth]{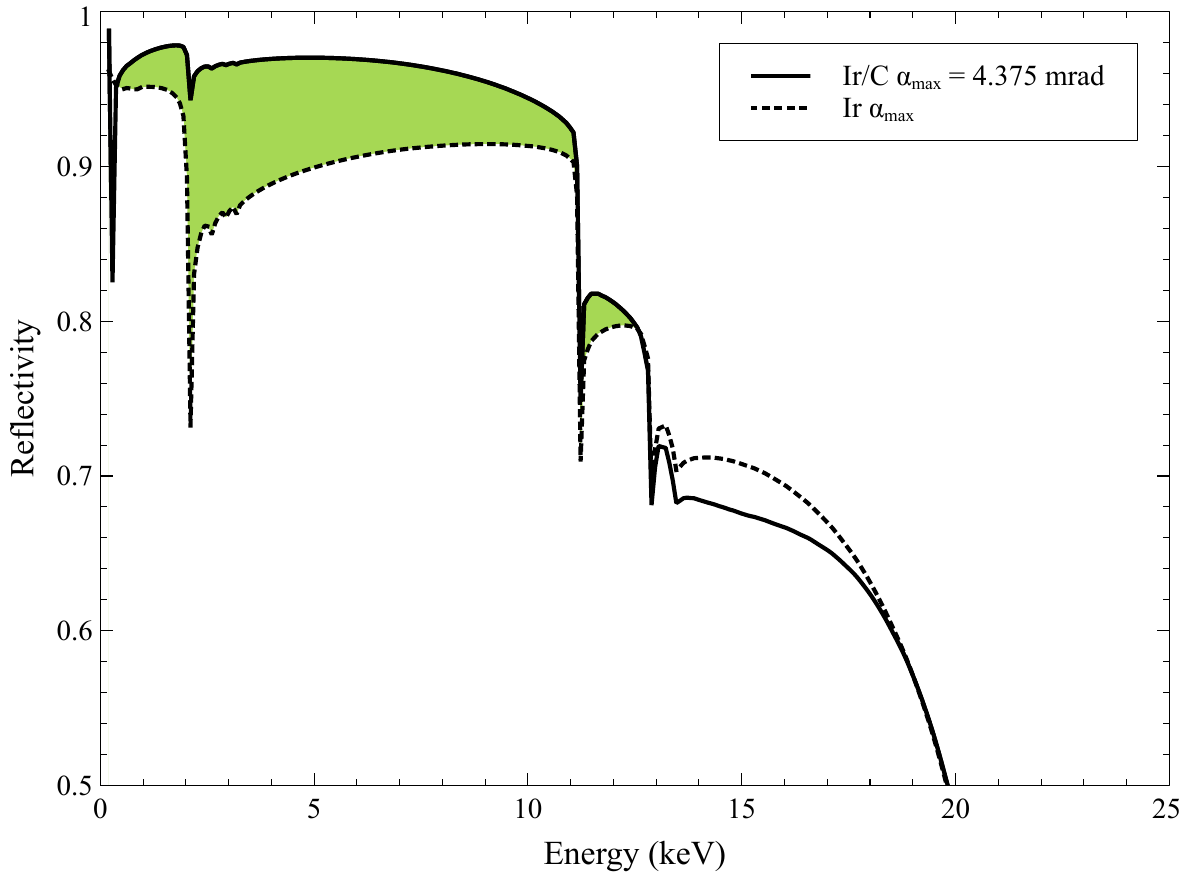}
    \caption{The C overcoat provides a significant improvement in reflectivity below 10~keV for a small trade between 15--20~keV. \textbf{Left}: Difference in reflectivity between a single Ir coated layer and Ir with a C overcoat for the minimum grazing incidence angle $\alpha_\mathrm{min} = 3.3$ mrad. \textbf{Right}: Difference in reflectivity between a single Ir coated layer and Ir with a C overcoat for the minimum grazing incidence angle $\alpha_\mathrm{max} = 4.375$ mrad.}
    \label{fig:ir_irc}
\end{figure}

The LEMA has been baselined with a 10~nm layer of Ir with an 8~nm C overcoat. This coating prescription is identical so what is proposed for Athena \citep{Svendsen2022}. The benefit of  the C overcoat is demonstrated in Figure~\ref{fig:ir_irc}: C mitigates  absorption occurring from the Ir M-complex, providing a significant improvement in low-energy performance.

Futher details on the mirror design and coating performance are presented in \cite{Madsen2023_1}.

\subsubsection{The Low Energy Detector Assembly: LEDA}
The LEDA utilizes one built-to-print detector quadrant of the WFI and packages it into a new housing. At the heart of the detector is the Si DEPFET sensor, which is a sensor-amplifier structure combined with a MOSFET integrated onto a fully depleted Si bulk. The DEPFET sensor is read out using two types of ASICs: the Switcher and the Veritas \citep{Herrmann2018}. These comprise the front-end electronics (FEE) of the DEPFET detector. The Switcher controls the row-steering of the DEPFET sensor in a rolling shutter mode by applying a sequence of voltages, where rows are read out sequentially, and all pixels of one row are read simultaneously. The Veritas is a low-noise, multichannel amplifier and shaper circuit with an analog output that serializes the analog signals of the DEPFET pixels, clocking out at a rate of 25.6 MHz. Each pair of Veritas/Switcher handles 64 rows of the DEPFET, and each detector contains 8 ASICs of each type.

\begin{figure}
    \centering
    \includegraphics[width=0.48\textwidth]{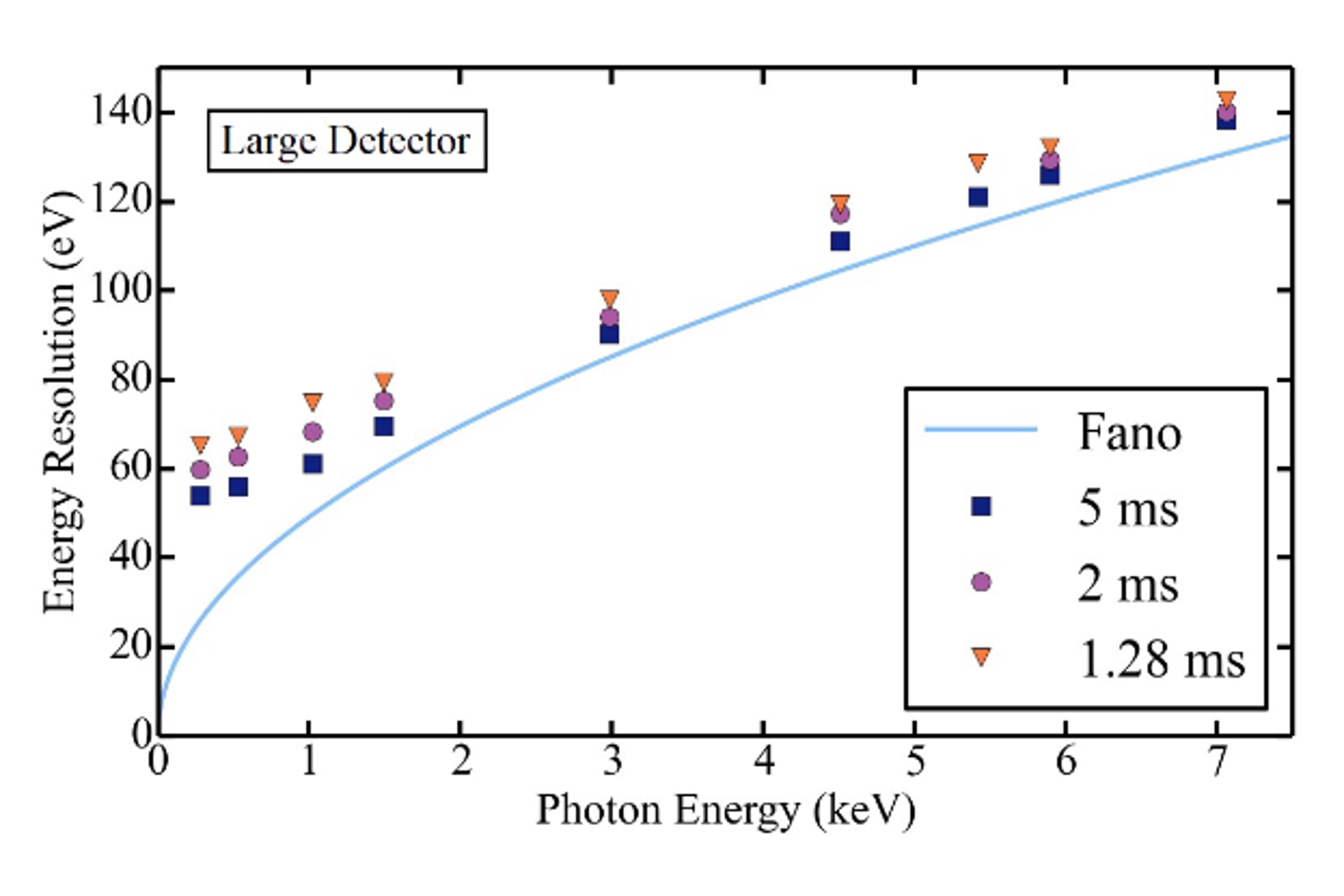}
    \includegraphics[width=0.46\textwidth]{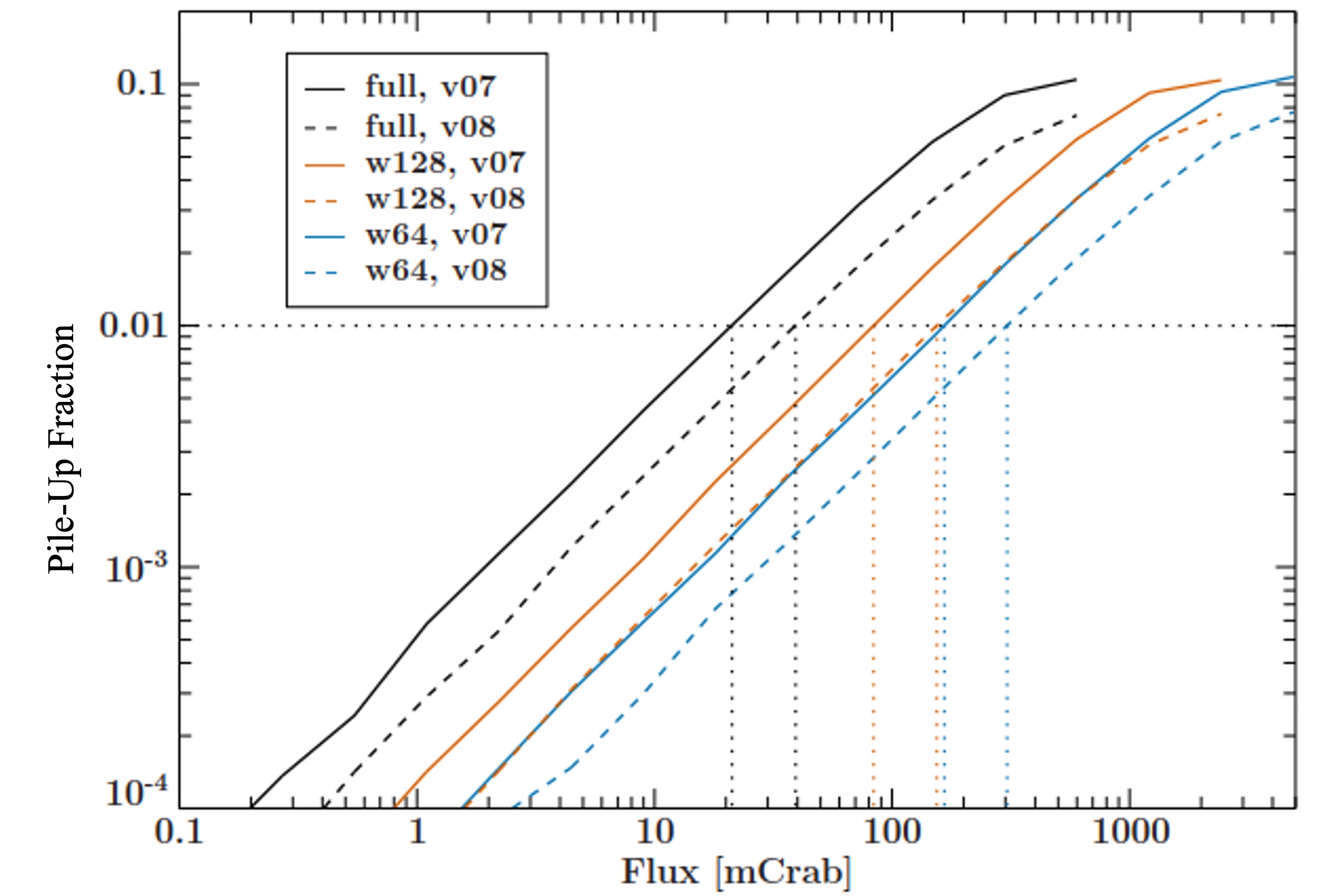}
    \caption{\textbf{Left}: Spectral performance of the DEPFET. The listed times are the exposure times, and therefore the time resolution of the sensor. For the readout, the full exposure time is used which allows for a better performance at a lower time resolution. The increasing difference between the theoretical Fano limit and the measured performance for lower photon energies is caused by charge losses at the entrance window. Low-energy photons have a lower penetration depth and increased probability of charge losses. \textbf{Right}: Simulated pile-up fraction for HEX-P in full frame, or window modes with either 64 (w64) or 128 (w128) rows.}
    \label{fig:leda}
\end{figure}

The DEPFET detector facilitates state-of-the-art spectral resolution over the required broad energy band from 0.2 to 20 keV, with a FWHM of 80 eV at 1~keV and 150~eV at 7~keV (Figure \ref{fig:leda}). High QE is achieved by back-illumination of the 450 $\mu$m thick sensor. A 5-$\mu$m thick benzocyclobutene (C$_8$H$_8$) layer on the off-mirror side is used for passivation and a 90-nm thick Al coating on the mirror side of the detector serves as an on-chip filter to reduce contamination of the X-ray signal by optical light. A filter wheel provides an additional optical blocking filter of 40~nm Al.

The LEDA is operated at 2~ms frame time (the WFI operates at 5~ms) in full frame mode (i.e., all pixels) at a small cost in energy resolution, as shown in Figure \ref{fig:leda} (left). For bright sources that cause pile-up, the detector can be operated in a window mode that enables a faster readout for fewer rows (in multiples of 64). For the smallest window (w64: 64 $\times$ 512 pixels), the detector readout is 250 $\mu$s. A Monte Carlo simulation using the SImulation of X-ray TElescopes (SIXTE) \citep{Dauser2019} software was used to estimate the LEDA pileup percentage as a function of incident count rate using a spectrum of the Crab with a slope of $\Gamma$=2.1. SIXTE contains the DEPFET simulator and can faithfully replicate the readout speed. Combined with the LET mirror effective area, SIXTE can predict the pileup fraction as a function of source flux. This was simulated for the full frame detector and for window modes using 128 or 64 rows. Figure \ref{fig:leda} (right) shows that 1\% pileup fraction is reached at 20 mCrab for the full detector and 150 mCrab for the 64-row mode. A higher flux limit could be allowed for a flatter spectrum.

Due to the potential for radiation damage from soft protons from the Sun, the LEDA has a magnetic diverter which deflects protons with energies $E <$ 75 keV \citep{Madsen2023_2}.

\subsubsection{Background}
The LEDA builds strongly on efforts made for Athena/WFI, and the radiation requirements on the shielding are directly inherited from the WFI \citep{Madsen2023_2}. The unfocused and focused CXB components add up to a mean rate of 1.0$\times 10^{-5}$ photons s$^{-1}$ cm$^{-2}$ keV$^{-1}$. The charged particle background, NCXB, is the dominant term and, based on the work done for the WFI, has a mean rate of 6.7$\times 10^{-3}$ photons s$^{-1}$ cm$^{-2}$ keV$^{-1}$. The Sun is a prompt source of soft X-rays and particles, which only irradiates the instrument for a short time before decaying. Since the periods of high background can be excluded, they are not included in the persistent instrument background estimate. The LET background components of the CXB and the charged particle NCXB are shown in Figure~\ref{fig:let_rsp} (left).

\subsubsection{Effective Area}
\begin{figure}
    \centering
    \includegraphics[width= 0.45\textwidth]{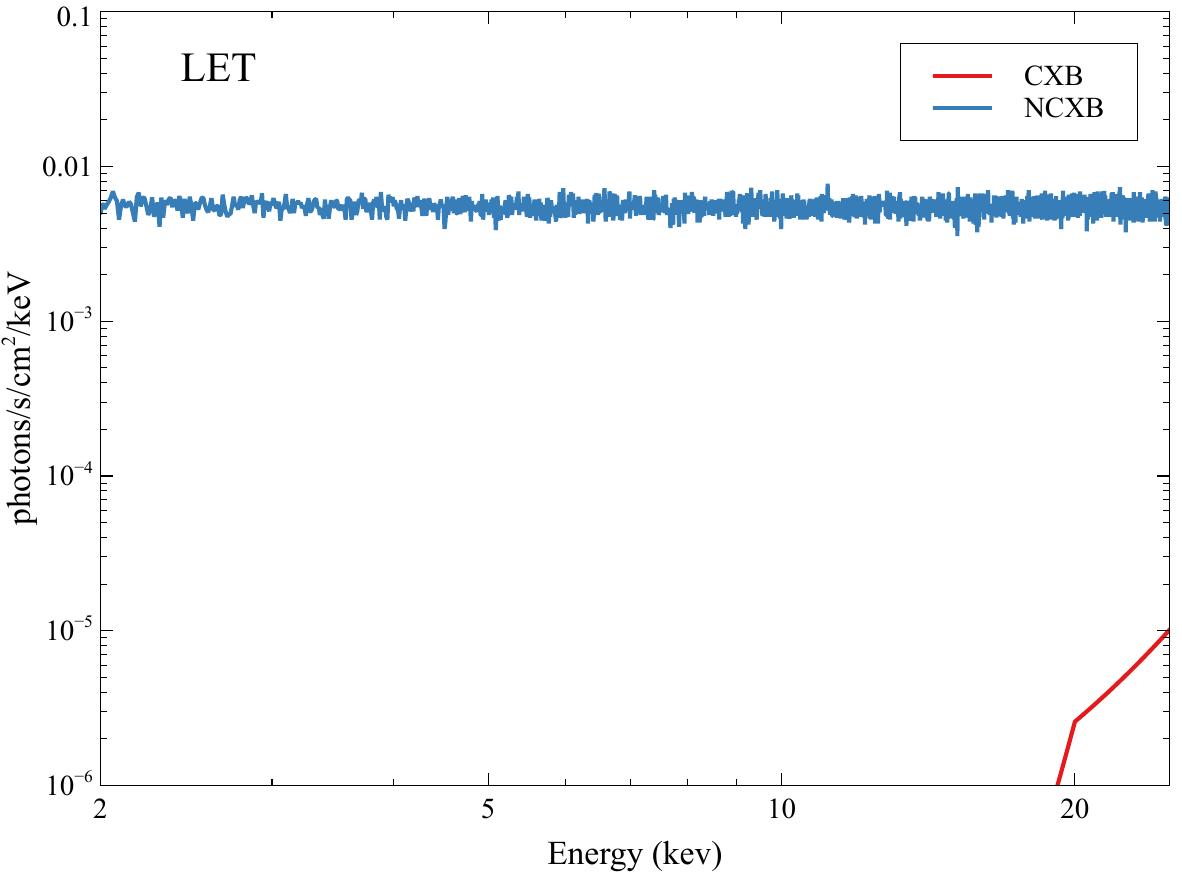}
    \includegraphics[width= 0.45\textwidth]{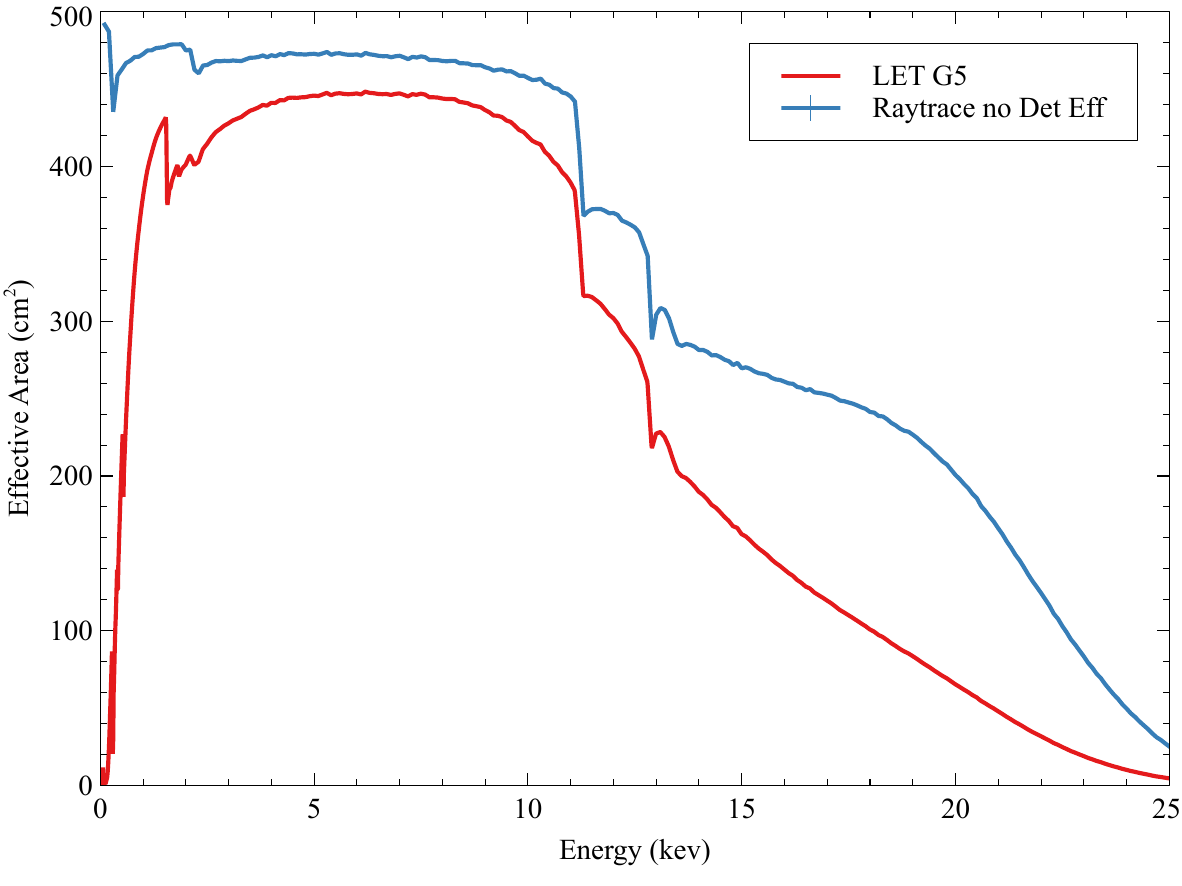}
    \caption{\textbf{Left}: The LET instrument background separated into the CXB and NCXB. \textbf{Right}: The effective area of the LET, with and without the detector efficiency and optical blocking filters.}
    \label{fig:let_rsp}
\end{figure}

The effective area of the LEMA was derived from a full ray-trace, including all known effects of obscuration, surface mirco-roughness, and multilayer reflection. The QE and redistribution matrix of the LEDA were simulated using the SIXTE detector simulator verified against laboratory data. Figure \ref{fig:let_rsp} (right) shows the best current estimate of the effective area of LEDA (v07) folded through the redistribution function of the LEDA detector and the optical blocking filters.

\section{Instrument Sensitivity}
Due to the broadband nature of the science objectives and the diversity of spectral features targeted in the source spectra, the HEX-P performance metric is expressed as the minimum detectable flux of a signal-to-noise ratio, $\sigma = S/\sqrt{S+B}$, where $S$ and $B$ are the source and background count rates in the bandpass of interest in the Poisson regime. This quadratic equation can be solved for $S$ to find the minimum count rate of the source for a given $\sigma$ and background rate $B$. Using the energy dependent instrument response functions, the count rate can be turned into a minimum detectable flux over a bin, $\Delta E$, by
\begin{equation}
    F_{min} = \frac{\sigma^2 + \sqrt{\sigma^4+4\sigma^2(btA_\mathrm{PSF}\Delta E)}}{2(A_\mathrm{Eff}t\Delta E)}
\end{equation}
where $b$ is the background flux (photons s$^{-1}$ cm$^{-2}$), $t$ is the exposure time, $A_\mathrm{PSF}$ is the area on the detector covered by the PSF, and $A_\mathrm{Eff}$ is the energy dependent effective area. The PSF detector area, $A_\mathrm{PSF}$, is optimized to maximize the signal in the background and has a radius of $\sim 1.2 \times $HPD. This flux can be expressed as: 1) the instrumental differential minimum detectable flux, $dF/dE$ (photon $^{-1}$ cm$^{-2}$ keV$^{-1}$), and 2) the minimum detectable flux in a specific bandpass, $F_{E1-E2}$ (erg s$^{-1}$ cm$^{-2}$), which gives the integrated flux of a specified spectrum between E1 and E2. 

\begin{figure}
    \centering
    \includegraphics[width=0.9\textwidth]{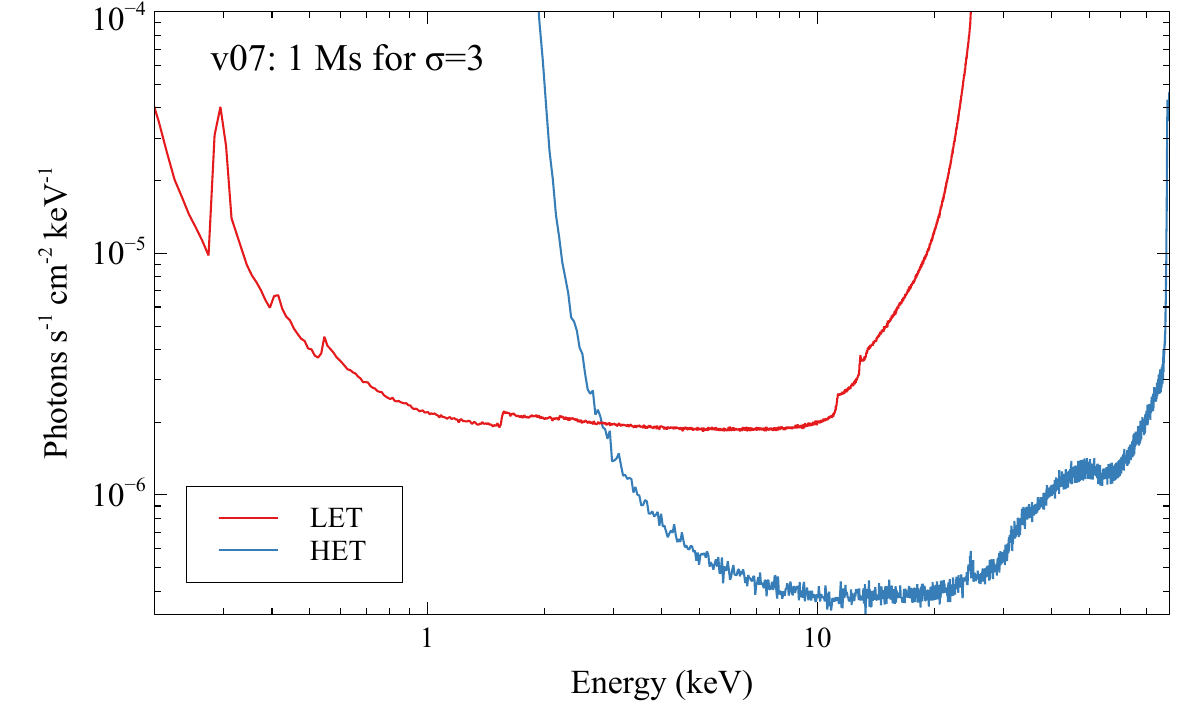}
    \caption{The differential flux, $dF/dE$, of the LET and HET based on the responses for the v07 instrument summarized in Table~\ref{tab:hexp}.}
    \label{fig:diff_flux}
\end{figure}

The differential flux of the instrument is shown in Figure \ref{fig:diff_flux}, and Table \ref{tab:sens} summarizes the sensitivity of the instrument for different bands, calculated for 1~Ms assuming an optimized signal-to-noise extraction region of 6\as\ (LET) and 12\as\ (HET) for a minimum detectable flux at $\sigma$ = 3 of a source with a power law photon index $\Gamma=1.9$.

\begin{table}[]
    \centering
    \caption{Sensitivity of the LET and HET.}
    \begin{tabular}{l|c||l|c}
    \hline
        \multicolumn{2}{c||}{HET} &  \multicolumn{2}{c}{LET} \\
        \hline
         Band & Sensitivity (erg s$^{-1}$ cm$^{-2}$) & Band & Sensitivity (erg s$^{-1}$ cm$^{-2}$) \\
         \hline
         2 - 10 keV & 1.1 $\times 10^{-15}$ & 0.2 - 5 keV & 3.6 $\times 10^{-16}$\\
         10 - 20 keV & 1.4 $\times 10^{-15}$ & 5.0 - 10 keV & 1.4 $\times 10^{-15}$\\
         20 - 40 keV & 5.2 $\times 10^{-15}$ & 0.5 - 10 keV & 6.0 $\times 10^{-16}$\\
         40 - 80 keV & 3.9 $\times 10^{-14}$ & 10 - 20 keV & 7.5 $\times 10^{-15}$\\
         \hline
         \multicolumn{4}{l}{Minimum detectable flux for $\Gamma=1.9$, $\sigma=3$, time = 1Ms}
    \end{tabular}
    \label{tab:sens}
\end{table}

\section{Mission Profile}
The HEX-P mission baseline is five years with 30\% of the observing time dedicated to the PI-led program and 70\% to the GO program. The spacecraft carries consumables for 20 years, and in the extended mission phase 100\% of the observing time will be dedicated to the GO program. During the baseline mission, the GO program is executed alongside the PI-led program.

The two instruments, HET and LET, are co-aligned and simultaneously observe the same field to provide the required broadband. At L1, the mission achieves an observing efficiency $\ge$ 90\% and an instantaneous $\le 3\pi$ field-of-regard for Target of Opportunity (ToO) observations. The observatory has a 24-hr on-target response time to ToOs, which is limited by the availability of uplink opportunities. The HEX-P Science Center (HSC) is located at Caltech, and ToOs will be evaluated by the science team and approved by the mission PI. The HSC will conduct an automated transient search on all data, sending out transient alerts through established systems.

The HSC manages the Science Data Center where all data undergo a quality assurance check before being sent to the High-Energy Astrophysics Science Archive Research Center (HEASARC) at GSFC to be archived and distributed to the community. Within two weeks of being taken, all science data are made publicly available without a proprietary period, in compliance with SPD-41a. 

The HSC will produce a serendipitous HEX-P Source Catalog, updated annually, to support the use of archival data. This catalog, based on the \xmm\, Serendipitous Survey \citep{Webb2020} and the \textit{Chandra} Source Catalog \citep{Evans2010}, will provide easy access to HEX-P observations with standardized data products. These can be used by a larger community less experienced in X-ray analysis and modeling. 

\section{Summary}
In summary, the HEX-P payload instrument has been designed to address the astrophysical questions highlighted by the 2020 Decadal Survey on Astronomy and Astrophysics. Instrument responses have been derived using advanced and detailed modeling of the instruments and the background, and they reflect the current best estimate of performance as of Spring 2023. The instrument responses have been made publicly available and in a format for use with XSPEC and SIXTE.

HEX-P will be launched into L1 with an observing efficiency $\ge 90$\%, and the 5-year mission profile has been designed to maximize the GO program with a 24-hour ToO response time. The PI-led and GO programs are executed in parallel and data are made available without a proprietary period.

\section*{Conflict of Interest Statement}

The authors declare that the research was conducted in the absence of commercial or financial relationships that could be construed as a potential conflict of interest.

\section*{Author Contributions}
KKM: Conceptualization, Formal Analysis, Investigation, Software, Supervision, Visualization, Writing–original draft, Writing–review and editing.
JG: Conceptualization, Writing–original draft, Writing–review and editing.
DS: Conceptualization, Writing–original draft, Writing–review and editing.
WZ: Investigation, Writing–review and editing.
SK: Formal Analysis, Investigation, Writing–original draft, Writing–review and editing.
AM: Writing–review and editing.
DS: Formal Analysis, Investigation, Software, Writing–original draft.
GP: Writing–review and editing.
SB: Formal Analysis, Investigation, Software, Writing–original draft.
BG: Formal Analysis, Investigation, Writing–review and editing.
KN: Writing–review and editing.
KP: Writing–review and editing
PP: Writing–original draft, Writing–review and editing.
DC: Formal Analysis, Investigation, Writing–review and editing.
AR: Formal Analysis, Investigation, Writing–review and editing.
JW: Software, Methodology, Writing – review and editing.
RS: Formal Analysis, Investigation, Writing–review and editing.


\section*{Acknowledgments}
The work of DS was carried out at the Jet Propulsion Laboratory, California Institute of Technology, under a contract with NASA.

\section*{Data Availability Statement}
The response files were produced for use with XSPEC and SIXTE and are made publicly available at \href{https://hexp.org/response-files}{https://hexp.org/response-files}.

\bibliographystyle{Frontiers-Harvard} 
\bibliography{bibliography}

\begin{thebibliography}{22}
\providecommand{\natexlab}[1]{#1}
\expandafter\ifx\csname urlstyle\endcsname\relax
  \providecommand{\doi}[1]{doi:\discretionary{}{}{}#1}\else
  \providecommand{\doi}{doi:\discretionary{}{}{}\begingroup \urlstyle{rm}\Url}\fi
\providecommand{\selectlanguage}[1]{\relax}
\providecommand{\bibAnnoteFile}[1]{%
  \IfFileExists{#1}{\begin{quotation}\noindent\textsc{Key:} #1\\
  \textsc{Annotation:}\ \input{#1}\end{quotation}}{}}
\providecommand{\bibAnnote}[2]{%
  \begin{quotation}\noindent\textsc{Key:} #1\\
  \textsc{Annotation:}\ #2\end{quotation}}

\bibitem[{Ayre et~al.(2018)Ayre, Bavdaz, Ferreira, Wille, Fransen, Stefanescu et~al.}]{Ayre2018}
Ayre, M., Bavdaz, M., Ferreira, I., Wille, E., Fransen, S., Stefanescu, A., et~al. (2018).
\newblock {ATHENA: system studies and optics accommodation}.
\newblock In \emph{Space Telescopes and Instrumentation 2018: Ultraviolet to Gamma Ray}, eds. J.-W.~A. den Herder, S.~Nikzad, and K.~Nakazawa. International Society for Optics and Photonics (SPIE), vol. 10699, 106991E.
\newblock \doi{10.1117/12.2313305}
\bibAnnoteFile{Ayre2018}

\bibitem[{Bavdaz et~al.(2021)Bavdaz, Wille, Ayre, Ferreira, Shortt, Fransen et~al.}]{Bavdaz2021}
Bavdaz, M., Wille, E., Ayre, M., Ferreira, I., Shortt, B., Fransen, S., et~al. (2021).
\newblock The athena x-ray optics development and accommodation.
\newblock 72.
\newblock \doi{10.1117/12.2599341}
\bibAnnoteFile{Bavdaz2021}

\bibitem[{{Brightman} et~al.(2023){Brightman}, {Margutti}, {Polzin}, {Jaodand}, {Hotokezaka}, {Alford} et~al.}]{Brightman2023}
{Brightman}, M., {Margutti}, R., {Polzin}, A., {Jaodand}, A., {Hotokezaka}, K., {Alford}, J. A.~J., et~al. (2023).
\newblock {The High Energy X-ray Probe (HEX-P): Sensitive broadband X-ray observations of transient phenomena in the 2030s}.
\newblock \emph{arXiv e-prints} , arXiv:2311.04856\doi{10.48550/arXiv.2311.04856}
\bibAnnoteFile{Brightman2023}

\bibitem[{Chen et~al.(2004)Chen, Cook, Harrison, Lin, Mao, and Schindler}]{Chen2004}
Chen, C., Cook, W., Harrison, F., Lin, J., Mao, P., and Schindler, S. (2004).
\newblock Characterization of the heft cdznte pixel detectors.
\newblock \emph{Proc SPIE} 5198.
\newblock \doi{10.1117/12.506075}
\bibAnnoteFile{Chen2004}

\bibitem[{Dauser et~al.(2019)Dauser, Falkner, Lorenz, Kirsch, Peille, Cucchetti et~al.}]{Dauser2019}
Dauser, T., Falkner, S., Lorenz, M., Kirsch, C., Peille, P., Cucchetti, E., et~al. (2019).
\newblock Sixte: a generic x-ray instrument simulation toolkit.
\newblock \emph{A\&A} 630, A66.
\newblock \doi{10.1051/0004-6361/201935978}
\bibAnnoteFile{Dauser2019}

\bibitem[{Evans et~al.(2010)Evans, Primini, Glotfelty, Anderson, Bonaventura, Chen et~al.}]{Evans2010}
Evans, I.~N., Primini, F.~A., Glotfelty, K.~J., Anderson, C.~S., Bonaventura, N.~R., Chen, J.~C., et~al. (2010).
\newblock The chandra source catalog.
\newblock \emph{The Astrophysical Journal Supplement Series} 189, 37.
\newblock \doi{10.1088/0067-0049/189/1/37}
\bibAnnoteFile{Evans2010}

\bibitem[{{Harrison} et~al.(2013){Harrison}, {Craig}, {Christensen}, {Hailey}, {Zhang}, and el~al.}]{Harrison2013}
{Harrison}, F.~A., {Craig}, W.~W., {Christensen}, F.~E., {Hailey}, C.~J., {Zhang}, and el~al. (2013).
\newblock {The Nuclear Spectroscopic Telescope Array (NuSTAR) High-energy X-Ray Mission}.
\newblock \emph{ApJ} 770, 103.
\newblock \doi{10.1088/0004-637X/770/2/103}
\bibAnnoteFile{Harrison2013}

\bibitem[{Harvey et~al.(2001)Harvey, Krywonos, Thompson, and Saha}]{Harvey2001}
Harvey, J.~E., Krywonos, A., Thompson, P.~L., and Saha, T.~T. (2001).
\newblock Grazing-incidence hyperboloid--hyperboloid designs for wide-field x-ray imaging applications.
\newblock \emph{Appl. Opt.} 40, 136--144.
\newblock \doi{10.1364/AO.40.000136}
\bibAnnoteFile{Harvey2001}

\bibitem[{Herrmann et~al.(2018)Herrmann, Koch, Obergassel, Treberspurg, Bonholzer, and Meidinger}]{Herrmann2018}
Herrmann, S., Koch, A., Obergassel, S., Treberspurg, W., Bonholzer, M., and Meidinger, N. (2018).
\newblock {VERITAS 2.2: a low noise source follower and drain current readout integrated circuit for the wide field imager on the Athena x-ray satellite}.
\newblock In \emph{High Energy, Optical, and Infrared Detectors for Astronomy VIII}, eds. A.~D. Holland and J.~Beletic. International Society for Optics and Photonics (SPIE), vol. 10709, 1070935.
\newblock \doi{10.1117/12.2314335}
\bibAnnoteFile{Herrmann2018}

\bibitem[{{Madsen} et~al.(2022){Madsen}, {Broadway}, and {Ferreira}}]{Madsen2022}
{Madsen}, K.~K., {Broadway}, D., and {Ferreira}, D. D.~M. (2022).
\newblock {Single-Layer and Multilayer Coatings for Astronomical X-Ray Mirrors}.
\newblock In \emph{Handbook of X-ray and Gamma-ray Astrophysics. Edited by Cosimo Bambi and Andrea Santangelo}. 71.
\newblock \doi{10.1007/978-981-16-4544-0_4-1}
\bibAnnoteFile{Madsen2022}

\bibitem[{{Madsen} et~al.(2004){Madsen}, {Christensen}, {Jensen}, {Ziegler}, {Craig}, {Gunderson} et~al.}]{Madsen2004}
{Madsen}, K.~K., {Christensen}, F.~E., {Jensen}, C.~P., {Ziegler}, E., {Craig}, W.~W., {Gunderson}, K.~S., et~al. (2004).
\newblock {X-ray study of W/Si multilayers for the HEFT hard x-ray telescope}.
\newblock In \emph{Optics for EUV, X-Ray, and Gamma-Ray Astronomy}, eds. O.~{Citterio} and S.~L. {O'Dell}. vol. 5168 of \emph{Proc. SPIE}, 41--52.
\newblock \doi{10.1117/12.505665}
\bibAnnoteFile{Madsen2004}

\bibitem[{Madsen et~al.(2023{\natexlab{a}})Madsen, Corsetti, Kenyon, Okajima, Rohrback, Tamura et~al.}]{Madsen2023_1}
Madsen, K.~K., Corsetti, J., Kenyon, S., Okajima, T., Rohrback, S., Tamura, K., et~al. (2023{\natexlab{a}}).
\newblock {Mirror design for the high-energy x-ray probe (HEX-P)}.
\newblock In \emph{Optics for EUV, X-Ray, and Gamma-Ray Astronomy XI}, eds. S.~L. O'Dell, J.~A. Gaskin, G.~Pareschi, and D.~Spiga. International Society for Optics and Photonics (SPIE), vol. 12679, 126791E.
\newblock \doi{10.1117/12.2677554}
\bibAnnoteFile{Madsen2023_1}

\bibitem[{Madsen et~al.(2023{\natexlab{b}})Madsen, Fleischhack, Violette, Grefenstette, Zoglauer, Arenberg et~al.}]{Madsen2023_2}
Madsen, K.~K., Fleischhack, H., Violette, D.~P., Grefenstette, B.~W., Zoglauer, A., Arenberg, J., et~al. (2023{\natexlab{b}}).
\newblock {Background simulations for the high energy x-ray probe (HEX-P)}.
\newblock In \emph{UV, X-Ray, and Gamma-Ray Space Instrumentation for Astronomy XXIII}, eds. O.~H. Siegmund and K.~Hoadley. International Society for Optics and Photonics (SPIE), vol. 12678, 126781B.
\newblock \doi{10.1117/12.2676891}
\bibAnnoteFile{Madsen2023_2}

\bibitem[{{Madsen} et~al.(2009){Madsen}, {Harrison}, {Mao}, {Christensen}, {Jensen}, {Brejnholt} et~al.}]{Madsen2009}
{Madsen}, K.~K., {Harrison}, F.~A., {Mao}, P.~H., {Christensen}, F.~E., {Jensen}, C.~P., {Brejnholt}, N., et~al. (2009).
\newblock {Optimizations of Pt/SiC and W/Si multilayers for the Nuclear Spectroscopic Telescope Array}.
\newblock In \emph{Optics for EUV, X-Ray, and Gamma-Ray Astronomy IV}. vol. 7437 of \emph{Proc. SPIE}.
\newblock \doi{10.1117/12.826669}
\bibAnnoteFile{Madsen2009}

\bibitem[{Meidinger et~al.(2020)Meidinger, Albrecht, Beitler, Bonholzer, Emberger, Frank et~al.}]{Meidinger2020}
Meidinger, N., Albrecht, S., Beitler, C., Bonholzer, M., Emberger, V., Frank, J., et~al. (2020).
\newblock {Development status of the wide field imager instrument for Athena}.
\newblock In \emph{Space Telescopes and Instrumentation 2020: Ultraviolet to Gamma Ray}, eds. J.-W.~A. den Herder, S.~Nikzad, and K.~Nakazawa. International Society for Optics and Photonics (SPIE), vol. 11444, 114440T.
\newblock \doi{10.1117/12.2560507}
\bibAnnoteFile{Meidinger2020}

\bibitem[{{Piotrowska} et~al.(2023){Piotrowska}, {Garc{\'\i}a}, {Walton}, {Beckmann}, {Stern}, {Ballantyne} et~al.}]{Piotrowska2023}
{Piotrowska}, J.~M., {Garc{\'\i}a}, J.~A., {Walton}, D.~J., {Beckmann}, R.~S., {Stern}, D., {Ballantyne}, D.~R., et~al. (2023).
\newblock {The High Energy X-ray Probe (HEX-P): Constraining Supermassive Black Hole Growth with Population Spin Measurements}.
\newblock \emph{arXiv e-prints} , arXiv:2311.04752\doi{10.48550/arXiv.2311.04752}
\bibAnnoteFile{Piotrowska2023}

\bibitem[{Riveros et~al.(2022)Riveros, Allgood, Biskach, DeVita, Hlinka, Kearney et~al.}]{Riveros2022}
Riveros, R.~E., Allgood, K.~D., Biskach, M.~P., DeVita, T.~A., Hlinka, M., Kearney, J.~D., et~al. (2022).
\newblock {Fabrication of lightweight silicon x-ray mirrors}.
\newblock In \emph{Space Telescopes and Instrumentation 2022: Ultraviolet to Gamma Ray}, eds. J.-W.~A. den Herder, S.~Nikzad, and K.~Nakazawa. International Society for Optics and Photonics (SPIE), vol. 12181, 1218111.
\newblock \doi{10.1117/12.2629017}
\bibAnnoteFile{Riveros2022}

\bibitem[{Saha and Zhang(2022)}]{Saha2022}
Saha, T.~T. and Zhang, W.~W. (2022).
\newblock Optical design of type-1 x-ray telescopes and their application to star-x.
\newblock \emph{Appl. Opt.} 61, 505--516.
\newblock \doi{10.1364/AO.446958}
\bibAnnoteFile{Saha2022}

\bibitem[{Svendsen et~al.(2022)Svendsen, Massahi, Ferreira, Gellert, Jegers, Christensen et~al.}]{Svendsen2022}
Svendsen, S., Massahi, S., Ferreira, D.~D., Gellert, N.~C., Jegers, A.~S., Christensen, F.~E., et~al. (2022).
\newblock {Characterisation of iridium and low-density bilayer coatings for the Athena optics}.
\newblock In \emph{Space Telescopes and Instrumentation 2022: Ultraviolet to Gamma Ray}, eds. J.-W.~A. den Herder, S.~Nikzad, and K.~Nakazawa. International Society for Optics and Photonics (SPIE), vol. 12181, 121810Z.
\newblock \doi{10.1117/12.2629976}
\bibAnnoteFile{Svendsen2022}

\bibitem[{{Vernani} et~al.(2022){Vernani}, {Bianucci}, {Grisoni}, {Marioni}, {Valsecchi}, {Keereman} et~al.}]{Vernani2022}
{Vernani}, D., {Bianucci}, G., {Grisoni}, G., {Marioni}, F., {Valsecchi}, G., {Keereman}, A., et~al. (2022).
\newblock {Follow-up x-ray telescope (FXT) mirror module for the Einstein probe mission}.
\newblock In \emph{Space Telescopes and Instrumentation 2022: Ultraviolet to Gamma Ray}, eds. J.-W.~A. {den Herder}, S.~{Nikzad}, and K.~{Nakazawa}. vol. 12181 of \emph{Society of Photo-Optical Instrumentation Engineers (SPIE) Conference Series}, 121811R.
\newblock \doi{10.1117/12.2630147}
\bibAnnoteFile{Vernani2022}

\bibitem[{Vernani et~al.(2011)Vernani, Borghi, Calegari, Castelnuovo, Citterio, Ferrario et~al.}]{Vernani2011}
Vernani, D., Borghi, G., Calegari, G., Castelnuovo, M., Citterio, O., Ferrario, I., et~al. (2011).
\newblock {Performance of a mirror shell replicated from a new flight quality mandrel for eROSITA mission}.
\newblock In \emph{Optics for EUV, X-Ray, and Gamma-Ray Astronomy V}, eds. S.~L. O'Dell and G.~Pareschi. International Society for Optics and Photonics (SPIE), vol. 8147, 814707.
\newblock \doi{10.1117/12.895329}
\bibAnnoteFile{Vernani2011}

\bibitem[{Webb et~al.(2020)Webb, Coriat, Traulsen, Ballet, Motch, Carrera et~al.}]{Webb2020}
Webb, N.~A., Coriat, M., Traulsen, I., Ballet, J., Motch, C., Carrera, F.~J., et~al. (2020).
\newblock Thexmm-newtonserendipitous survey: Ix. the fourthxmm-newtonserendipitous source catalogue.
\newblock \emph{Astronomy and Astrophysics} 641, A136.
\newblock \doi{10.1051/0004-6361/201937353}
\bibAnnoteFile{Webb2020}

\end{thebibliography}


\end{document}